\documentstyle[aas2pp4]{article}

\received{15 June 1999}

\slugcomment{The Astrophysical Journal, Submitted 2001 March 23}

\input{psfig}
\begin{document}

\title{$r$-PROCESS IN PROMPT SUPERNOVA EXPLOSIONS REVISITED}

\author{\sc K. Sumiyoshi$^{1}$,
             M. Terasawa$^{2,3,4}$,
             G. J. Mathews$^{3,5}$,
             T. Kajino$^{2,3,6}$
             S. Yamada$^{7}$,
             and H. Suzuki$^{8}$
}
\affil{\em sumi@la.numazu-ct.ac.jp}
%
\bigskip
\affil{The Astrophysical Journal, Submitted 2001 March 23}

\altaffiltext{1}{Numazu College of Technology (NCT),
Ooka 3600, Numazu, Shizuoka 410-8501, Japan}
\altaffiltext{2}{Department of Astronomy, The University of Tokyo,
                  Hongo 7-3-1, Bunkyo, Tokyo 113-0033, Japan}
\altaffiltext{3}{National Astronomical Observatory,
                  Osawa 2-21-1, Mitaka, Tokyo 181-8588, Japan}
\altaffiltext{4}{Institute of Physical and Chemical Research (RIKEN)
                  Hirosawa 2-1, Wako, Saitama 351-0198, Japan}
\altaffiltext{5}{Center for Astrophysics, Department of Physics,
                  University of Notre Dame, Notre Dame, Indiana 46556, USA}
\altaffiltext{6}{The Graduate University for Advanced Studies,
                  Osawa 2-21-1, Mitaka, Tokyo 181-8588, Japan}
\altaffiltext{7}{Institute of Laser Engineering (ILE), Osaka University,
                  Yamadaoka 2-6, Suita, Osaka 565-0871, Japan}
\altaffiltext{8}{Department of Physics, Science University of Tokyo,
  Yamazaki 2641, Noda, Chiba 278-8510, Japan}

\begin{abstract}
We reanalyze $r$-process nucleosynthesis in the neutron-rich
ejecta from a prompt supernova explosion of a low-mass
(11 M$_\odot$) progenitor.  
Although it has not yet been established
that a prompt explosion can occur, it is not yet ruled out
as a possibility for low-mass supernova progenitors.
Moreover, there is mounting evidence that a new $r$-process
site may be required.  Hence, we assume that a prompt explosion can occur
and make a  study of r-process nucleosynthesis 
in the supernova ejecta.  To achieve a prompt explosion 
we have performed a
general relativistic hydrodynamic simulation
of adiabatic collapse and bounce using a
relativistic nuclear-matter equation of state.
The electron fraction $Y_e$ during the collapse 
was fixed at the initial-model value.
The size of the inner collapsing core
was then large enough to enable a prompt
explosion to occur in the hydrodynamical calculation.
Adopting the calculated trajectories of promptly ejected material,
we explicitly computed the burst of neutronization
due to electron captures on free protons
in the photodissociated ejecta after the passage of the shock.
The thermal and compositional evolution of
the resulting neutron-rich ejecta originating from 
near the surface of proto-neutron star was obtained.
These were used in nuclear reaction network calculations
to evaluate the products of $r$-process nucleosynthesis.
  We find that, unlike earlier studies of nucleosynthesis in prompt
supernovae, the amount of $r$-process
  material ejected per supernova
is quite consistent with observed galactic $r$-process abundances.
Furthermore, the computed $r$-process abundances are in good agreement with
Solar abundances of $r$-process elements for A$> 100$.
This suggests that prompt supernovae are still a viable $r$-process
site.  Such events may be responsible
for the abundances of the heaviest $r$-process nuclei.
\end{abstract}


\section{Introduction}

The astrophysical site for the origin of elements by
rapid neutron capture (the $r$-process) has been a mystery for some time.
Numerous sites have been proposed as possible candidate environments
(Hillebrandt 1978; Mathews \& Ward 1985; 
Mathews \& Cowan 1990; Meyer 1994).
At the present time the most popular
among proposed sites involves
the neutrino-heated ejecta from a nascent neutron
star (\cite{Woos92}; \cite{Meyer92}; \cite{Woos94}; \cite{otsuki};
\cite{Sum00}; \cite{Ter01}).
In  Woosley et al.~(1994) it was  demonstrated
that the Solar $r$-process abundances
were well reproduced in material which has been ablated from the
proto-neutron star in neutrino-driven winds.

There are,
however, a few serious problems with this $r$-process paradigm. For one,
elements with $A \sim 90$ are significantly overproduced by
over a factor of $\sim 100$.
Second, the requisite high entropy ($S/k \gtrsim 400 $) in
the supernova simulations has not been duplicated by other
independent theoretical studies (cf. \cite{Witt94}; \cite{Taka94};
\cite{Qian96}).  A third problem (\cite{Meyer98}) is the possibility
that neutrino-nucleus interactions could drastically alter the
neutron to seed ratio rendering a satisfactory $r$-process more difficult.

On the other hand, other viable sites have been
demonstrated to also account for the Solar $r$-process abundance pattern.
For  example,  it has been shown (\cite{Frei99b}) that neutron-star
mergers can produce the abundances of nuclei with  $A \gtrsim 130$.
Collapsing O-Ne-Mg cores resulting from progenitor stars
of low mass  $\sim 10 M_\odot$ also remain (\cite{Whee98})
as a promising site for the $r$-process.
  The present work, however, is
primarily concerned with collapsing iron cores of
$\sim 11 M_\odot$ progenitor stars.

   In several early papers  (e.g.~Hillebrandt et al.~1976;
Hillebrandt 1982;
Hillebrandt et al.~1984)
studies were made of $r$-process nucleosynthesis in the
material hydrodynamically ejected during a prompt supernova explosion.
In Hillebrandt et al.~(1976)
the $r$-process was thought to occur in the inner  $\sim 0.43$ M$_\odot$
of ejected material in which
the neutron to proton ratio  was taken to range
from  $n/p \sim 1$ to $7$.
The problem with these earlier studies, however,
is that it appeared that too much material was ejected
to make a reasonable accounting for the galactic abundances of
$r$-process nuclides.  To account for the current galactic abundance
of $r$-process material requires that only about $10^{-4}$ M$_\odot$
of $r$-process elements be ejected per supernova
(cf. Mathews \& Cowan 1990; Woosley et al.~1994).
This is much less than the ($\sim 0.4$ M$_\odot$) ejected
in the Hillebrandt et al.~(1976) model.    
Furthermore, it gradually became clear 
(\cite{Bar90,Bethe90,Suz94}) that
models of prompt supernovae could only 
be made to explode for a limited
range of small core-masses, high lepton fractions, and equation of state
parameters, and may not occur at all. 
Because of these dilemmas,
prompt supernovae have fallen out of favor as an
$r$-process paradigm in recent years.

  However, recent studies (\cite{Heger01}) of progenitor cores based 
upon improved  slower electron capture rates (\cite{Lang00}) 
have shown that smaller, cooler, iron cores
with larger lepton fractions are a likely outcome of supernova 
progenitor-star evolution.  
Although it has not yet been established that a prompt explosion 
can occur, the newer electron capture rates
may suggest that they are possible.

The purpose
of the present work is, therefore,  to revisit the  question of whether
prompt supernovae necessarily overproduce the abundances
of $r$-process elements.
We assume that a prompt
explosion can occur under the plausible conditions described below.
We then  examine the associated r-process nucleosynthesis.
Our purpose is not to provide a fully detailed prompt explosion model,
which will take some time to complete.  Instead, our purpose here
is merely to develop a basic prompt explosion model with enough
content to examine the question of the implied $r$-process
yield and relative abundances.  Therefore,
we begin with an adiabatic collapse simulation with the
electron fractions ($Y_e$) of the various zones fixed 
at the initial-model values.
This approximates the effect of the new diminished electron
capture rates.  This model produces a prompt explosion,
both because the core bounce is energetic, and because the
higher $Y_e$ leads to a small 
outer iron core which does not have enough material to dissipate
the shock by photodissociation.    It is adequate for our purpose.

A key part of the $r$-process calculation of the present
work is that we perform post processing of the 
compositional change of the ejecta relevant to determining the amount
of $r$-process material ejected.
First, we consider the burst of neutronization
due to electron capture on free protons
after the shock heating and photodissociation of the ejecta.
We show that the neutronization is rapid.  Moreover, the amount
of neutronized ejecta is much less than was assumed in earlier work.
Consequently, the correct amount of neutron rich material is ejected
to account for galactic abundances.  Our network calculations
of the ensuing $r$-process nucleosynthesis makes a reasonable
reproduction of the Solar $r$-process abundance distribution
for nuclei with $A \ga 100$.  This suggests that prompt supernovae 
remain as a viable candidate site for the $r$-process 
synthesis of heavy elements.

\section{Prompt Supernova Mechanism}

   For some time it has been debated in the literature as to whether
collapsing iron cores of supernova progenitors
can explode via a prompt mechanism (\cite{Bethe90,Suz94}).  
It has not yet been established that a prompt explosion can occur.
The question has been whether
the core bounce itself can be sufficiently energetic
to eject the outer layers of the star.   It is always difficult
to develop a supernova model which actually explodes 
(\cite{Bethe90,Suz94}) and
it has been clear for some time that if a prompt explosion
is to occur it must involve a low-mass ($\sim 10-12$ M$_\odot$)
progenitor star and a low-mass  collapsing
core (e.g. \cite{Bar90}).  Whether or not an explosion can occur 
depends upon a difference
between two large numbers: the energy in the outward moving
shock and the energy lost by the shock due to the photodissociation
of nuclei.  At present there are too many uncertainties in the
input supernova physics to definitively
exclude a prompt explosion mechanism for at least a limited 
range of core masses and equation of state parameters.  
Nevertheless, it is clear what ingredients
are necessary.  Two key factors which we invoke here to
obtain a prompt explosion are a small total iron core mass
and a large $Y_e$ as a result of slower electron capture
rates.  The important roles
of these we now outline in a brief summary of the explosion
mechanism.

    As the electron-degenerate iron core
reaches its Chandrasekhar mass, the collapse divides into an
inner homologously collapsing core and an outer core collapsing more slowly.
A prompt explosion requires 
a small outer iron core so that the outward moving shock from
the bounce of the inner core is not dissipated by photodissociation of
the outer core.  A small outer core can be achieved by minimizing the
total Chandrasekhar mass of the iron core, while
maximizing the mass of the inner homologous core.
The  Chandrasekhar mass is smallest for the cooler cores of the lowest-mass 
supernova progenitors.
A sufficiently small, cool,  iron core  ($M_{core} \sim 1.3$ M$_\odot$) 
results from the 11 M$_\odot$ progenitor star model 
 of (\cite{WW95})  employed in the present work.
 
A large inner  homologous core can result
if  the lepton fraction 
is large (\cite{Tak84,Yam94}).  This can happen if the
electron capture rates are slower than the usually employed
rates of Fuller, Fowler, \& Newman (1980;1982ab;1985) [hereafter FFN].
Electron capture rates are important because
degenerate electrons account for a large fraction of the
pressure support of the inner core.
A larger $Y_e$ therefore implies more pressure support, and 
a larger inner core.  
Ultimately, during the collapse, the neutrinos become trapped.
The  electron capture rates are then suppressed by the high neutrino
Fermi energy which limits the available  phase space.  
Afterward, the lepton fraction remains fixed. 

The collapse of the  inner core  slows as the central density exceeds
nuclear density and an outward moving shock is produced.
A somewhat soft equation of state  helps
the explosion (\cite{Bar90}), but it is not as important as
keeping the lepton fraction  large.

If the outer part of the core is small enough, it will not
completely absorb the outward going shock
by the photodissociation of iron-group nuclei.
The shock is then not dissipated.  It eventually reaches
the surface of the outer iron core and an explosion can ensue.
Therefore, this combination of conditions (i.e.~a large lepton
fraction and  a small total iron core mass) enables energetic,
prompt explosions and the ejection of neutronized material
from deep inside the star.

If the standard FFN electron capture rates are employed
even smaller iron cores  ($M \approx 1.2$ M$_\odot$)
cannot explode promptly (\cite{Rampp00,Lieb01}; cf. \cite{Bar90}).
 With these rates, the lepton
fraction is too small and so is the inner core.
However, with slower electron capture rates,
a prompt explosion is possible.
The determination of electron capture rates in supernova
cores has, however, suffered from large uncertainties in
the associated nuclear physics.  Moreover, 
recent studies (e.g. \cite{Lang00}) of electron capture on neutron-rich nuclei
have reported rates which are on average an order of magnitude smaller
than the previous standard rates of FFN.  Indeed,
a recent study (\cite{Heger01}) of the effect of these rates on
15-40 M$_\odot$ progenitor stars found significantly
larger $Y_e$ and cooler, smaller, cores.
Electron capture by free protons in a collapsing
core can also be suppressed if the free proton fraction
is small.  A small proton fraction can result from
a large symmetry energy in the
equation of state.  This is the case
for the one we adopt in the current
study (\cite{Shen98a,Shen98b}).

Clearly, further numerical simulations of the formation and collapse of 
the small iron cores associated with 
improved electron capture rates and a good equation of state are
called for.  This, however, is beyond the scope of
the present work.  Nevertheless, the model described here is
a plausible version of the required physics in order
to study the $r$-process elements produced.

\section{Explosion Model}
In the current study, we focus on the consequences
for r-process nucleosynthesis of a prompt supernova explosion.
In order to investigate mass ejection
in a prompt explosion  an adiabatic hydrodynamic collapse
calculation was performed.   As an extreme case, we assume that
the electron captures are sufficiently slower than the standard FFN rates
that the electron fraction remains fixed during the collapse.
This maximizes the shock energy, and therefore,
the explosion energy,  and brings the iron
core to a prompt explosion.
We use this hydrodynamical calculation to 
evaluate the amount of neutron-rich material
ejected and the thermodynamic conditions
during nucleosynthesis.

Our ultimate goal is a full calculation of
the  neutrino-radiation hydrodynamics.
It is, however, computationally difficult to follow this evolution
through the long time scale over which the nucleosynthesis occurs.
The work described below is a necessary
first preliminary step toward that
goal.

\subsection{Equation of State}
One important ingredient for a calculation of a prompt explosion
is the nuclear equation of state.
Fortunately, recent information on the physics of unstable nuclei
is helping to clarify both supernova explosions and the r-process.
One can, for example, now
use radioactive nuclear beam facilities to
  probe both the neutron-rich matter of supernovae and the nuclear data
needed for r-process nucleosynthesis calculations.
Having this experimental information on unstable nuclei
with large asymmetry,
Sumiyoshi et al.~(1993; 1995a; 1995b) and Hirata et al.~(1997)
  have extensively studied
nuclear structure and nuclear matter within a
relativistic many-body framework.
Based upon this work a relativistic EOS table
for application   to supernova simulations (\cite{Shen98a,Shen98b})
has recently been developed.
This relativistic EOS table enables one to perform full
simulations of supernovae from the initial gravitational core
collapse to the ejection and cooling of material from
the newly formed neutron star.
It has successfully been applied to studies of
neutrino-driven winds  from proto-neutron stars (\cite{Sum00}) and
numerical simulations of core collapse (\cite{Sum01}).
Although this equation of state is not particularly soft (incompressibility
$=$281 MeV), it does, have a large symmetry energy (36.9 MeV).
Since nuclei with large asymmetries were employed in the
development of this equation of state, one expects that the symmetry
energy is better determined.  The large deduced value  
reduces the free proton fraction.
This can raise the $Y_e$ for the core since there are fewer protons
available for electron capture.

\subsection{Hydrodynamics}
  The prompt supernova calculations described herein are based upon the
general-relativistic spherically-symmetric hydrodynamics code of Yamada (1997).
It makes use of a fully implicit Lagrangian numerical scheme.
It has been designed for the study of
supernova explosions and treats both the hydrodynamics and the
neutrino transport.

Although a  Boltzmann solver for the neutrino transport has recently
been implemented (Yamada, Janka, \& Suzuki 1999)
and detailed simulations of gravitational core collapse are now
being undertaken, in the current study we have performed
calculations of core collapse using
only pure hydrodynamics without neutrino transport.
Since we are primarily interested here in
the dynamics of collapse and prompt ejection,
we can as a first approximation ignore the
late time neutrino diffusion and heating
which is not manifest till later times in the explosion.
Indeed, the present work
is mostly concerned with matter ejected within
$\sim 0.2$ sec after core bounce and largely before the
arrival of the delayed neutrinos.
Obviously, this is an approximation which we will
address in a subsequent work.

The present study, however,  should be adequate
to analyze the physics of the hydrodynamics
of core collapse, bounce and a possible explosion
with our adopted EOS.
In particular, our
primary purpose here is to evaluate the amount of ejected
material which contributes to the r-process.
For this particular purpose, our neglect of
neutrino  radiative transport is probably adequate.

For the same reasons, we  assume that the collapse is adiabatic
collapse. That is  we neglect
the heating and cooling via neutrinos.
Thus, the entropy per baryon of each mass element
remains constant during the collapse except for during
the passage of the shock.

Altogether, these approximations tend to
maximize the energy of the prompt explosion
and provide an upper bound on the amount of ejecta.
Thus, our estimates of ejected $r$-process material can perhaps
be taken as an upper limit which is what we seek in the present study.

\subsection{Presupernova Model}
We adopt the presupernova model of a 
11 $M_{\odot}$ progenitor as computed
by Woosley \& Weaver (1995).
This is the best case for a prompt explosion
as it provides 
the smallest iron core ($M_{core} = 1.32$ M$_{\odot}$). It is also
the smallest progenitor mass
among their models ranging from 11$M_{\odot}$ to 40$M_{\odot}$.
 Both of these conditions are favorable for a prompt explosion.
For our hydrodynamic simulations we are only concerned
with the central 1.62 $M_{\odot}$,
which contains the iron core and outer layers.
We adopt densities, electron fractions, temperatures
and radii as a function of the baryon-mass mesh
from the progenitor model.  Other quantities
are derived using our adopted
relativistic EOS table.

The simulations utilize 200 baryon-mass mesh points.
This provides enough resolution to analyze
the $r$-process in the ejecta.
The inner 50 grid points span a range of baryon mass from 0 up to 1.0
$M_{\odot}$.
The outer 150 grid points span
the remaining 0.62 $M_{\odot}$.
The grid size is taken to be  uniform in each of these two regions.
We have checked our results in simulations using different mass meshes
and find that they do not appreciably alter the nucleosynthesis results.

For the outer boundary conditions, we  have fixed quantities
at the values obtained at the cut position in the 11 M$_\odot$
progenitor model.

\section{Results}
\subsection{Collapse and Shock Ejection}

The central core
of the initial presupernova model is already marginally
gravitationally  unstable. Therefore,
when we start the numerical calculation with our
adopted initial conditions, the inner core starts collapsing
immediately.
Because we fix $Y_{e}$ as described in the next
subsection, the bounced inner core size is large enough for
the shock to realize a prompt explosion with matter ejection.

Figure 1 shows the evolution of various mass shells
as a function of time during the collapse.   Here we see that
a well formed outward going shock develops within $\sim 100$ ms
after core bounce.  This short time scale justifies our neglect
of neutrino transport
in this calculation, since delayed neutrinos would not be expected to
affect the explosion till several hundred milliseconds after core bounce.

We have extracted trajectories from the numerical
prompt explosion results
for use in the subsequent
calculations of electron capture on protons
and r-process nucleosynthesis.
The trajectories with grid numbers ranging from 90 to 111
out of 200 grid points are responsible for the $r$-process.
These are used for the nucleosynthesis
calculations. The main features of these trajectories are summarized in
Figures 2a-2d.

 From Figures 2a and 2b we see that these mass shells are compressed
during the bounce to densities in excess of  $10^{11}$ g cm$^{-3}$ and
are heated to temperatures well above $\sim 1$ MeV due to the
passage of the shock.
This heating is more than sufficient to  completely photodissociate this
material into  free neutrons and protons.  Electron capture on the
free protons then leads to a
burst of neutronization.

Figure 2c  shows the entropy per baryon in the outgoing ejecta.
The obtained entropy ($S/k \sim 10$) is much less than that
obtained in supernova wind models (e.g.~Woosley et al.~1994),
but it is still sufficiently high that heating by the subsequent $r$-process
nucleosynthesis  (Sato 1974; Hillebrandt et al. 1976)
does not appreciably affect the hydrodynamic evolution
of the ejecta.  To see this,
consider that the total Q-value of a typical
  beta-decay chain back to stability (e.g. $^{130}$Cd $\rightarrow$
$^{130}$Te) is about 20 MeV (i.e. $\sim 3\times 10^{-5}$ erg per $r$-process
isotope produced).  The ejected mass is 0.025 M$_\odot$
or  $\sim 3 \times 10^{53}$  nuclei (assuming A $\sim 100$).  Therefore,
the total energy released is only $9 \times 10^{48}$  erg, which is
only $\sim 1\%$ of the explosion energy.
In earlier studies of $r$-process heating (Sato 1974; Hillebrandt et al. 1976)
the entropy was much smaller, $\sim 0.1 - 1$, so that heating by the
$r$-process
was much more important for the dynamics of the ejecta.

\subsection{Electron Fraction in the Ejecta}

During the core collapse, we assume complete suppression of
electron capture on neutron-rich nuclei and free protons
in the supernova core.
We fix $Y_e$ to the original values of the progenitor core
during the hydrodynamical calculation of the core collapse.
(We calculate, however, compositional change after
the shock passage separately, to be discussed below.)
This assumption overestimates the electron fraction
(or lepton fraction after the neutrino trapping)
during the collapse.
It provides us with a large inner core, and hence,
enough shock energy for a successful prompt explosion.
We expect that this approximation might lead
to an overestimate of the amount of ejected $r$-process
material.  If that is the case, then it is adequate for our purpose,
which is to demonstrate that not too much
$r$-process material is ejected.

After the core bounce, the shock passes through the
outer part of the iron core and dissociates iron-group
nuclei into free neutrons and protons.
Near to the remnant, the density reaches $\sim 10^{11}$ g cm$^{-3}$.
Here, electron capture on free protons occurs abruptly
because of the high ambient electron degeneracy.
The material becomes immediately transparent to neutrinos
due to the dissociation of the nuclei which had been the cause of
neutrino trapping.
Neutrino absorption, which is the reverse of
the electron capture, is negligible.   This process
does not depend upon the details of neutrino transport.
It corresponds to the well known
 neutronization burst in the flux of
supernova neutrinos.  It causes the material to quickly neutronize
right after the shock passes. 
The resulting neutron-rich region can be generally seen in many supernova
simulations (e.g. \cite{Suz94}).  The inner regions near the
nascent neutron star tend 
to be neutron-rich.
It is this electron capture process in the ejecta that produces the neutron
excess necessary to achieve r-process conditions.

To treat this neutronization,
we have calculated the time evolution
of the electron fraction  $Y_e$ due to electron capture on
free protons by postprocessing the results of
the hydrodynamical calculation.
This electron capture drastically reduces the electron
fraction and then ceases quickly due to
the rapid expansion of the ejecta.
The degree of neutronization depends on the compression
of matter at the bounce.
The general tendency is that
material deeper in the ejecta has a higher electron
degeneracy and a smaller electron fraction.
The resulting electron fractions
in various ejected mass shells then become initial conditions
for our subsequent r-process nucleosynthesis calculations
(see Figure 2d).


We should point out that this explicit determination of the
neutronization of the ejecta is an improvement over the
previous hydrodynamic calculations of Hillebrandt et al.~(1976).
In that work a continuous distribution in $Y_e$ between $Y_e = 0.5$ at
the base of the silicon-burning shell and $Y_e = 0.125$ at
the remnant neutron star was
simply assumed for the $r$-process initial conditions.
As we shall see, our calculations indicate that a much smaller
region of neutronized material actually emerges from
a prompt explosion.

  Figure 3  summarizes the electron fraction
of material at the beginning of nucleosynthesis
as a function of baryon mass coordinate in the ejecta.
The dotted line displays the initial value of $Y_{e}$ before the
neutronization.
As we shall see the $r$-process occurs only in regions with
$Y_e \sim 0.15$ to $0.25$.  In this calculation that
corresponds to only 0.025 M$_\odot$ of material.  This is much less than
in the earlier estimates of Hillebrandt et al (1976).
The difference between the present results and their calculation
can be traced to their simplified treatment of the neutronization
of the ejecta.
We stress again that we have performed hydrodynamical calculations
to find the ejected part of material and have explicitly
calculated the electron fractions of each mass elements
due to electron captures on free protons in order to determine
both the amount and neutron-richness of ejecta.

The grid point 90 ($M_{b}=1.163 M_{\odot}$ at zone center in baryon
mass coordinate) is the inner-most  mass shell of
the ejected material.
This trajectory has the smallest electron
fraction ($Y_e = 0.16$ in Figure 3).
It contributes the most to
the $r$-process elements around  the A=195 abundance peak.
The grid points from 90 to 95 ($M_{b}=1.184 M_{\odot}$),
leading to $Y_e \la 0.25$
contribute to the r-process nucleosynthesis
from A=130 to A=190.
The grid point 111 ($M_{b}=1.250 M_{\odot}$)
has an electron fraction 0.45.
This trajectory only produces elements up to $A=70$
and does not contribute to the $r$-process.
Accordingly, we do not add outer mass shells (grid points $>112$)
in the following nucleosynthesis calculations as they do not
contribute to the abundances of $r$-process elements.

\section{$r$-Process Nucleosynthesis}

As the neutronized material is ejected it quickly experiences
charged-particle and neutron-capture reactions to produce seed material
followed by neutron captures and beta decays along the $r$-process path.
   The $r$-process nucleosynthesis calculations in the present work
are based upon the nuclear reaction network as described in Terasawa
et al.~(2001).  In that work it was shown that both
neutron-rich light-mass nuclei as well as heavy nuclei can play an
important role in the production of both seed nuclei and $r$-process
elements. The path through light nuclei is most important in
very neutron rich conditions such as in the high entropy
neutrino-driven wind or in the most neutron-rich ejecta studied here.
We  thus make use of an extended nuclear reaction network which includes
all relevant nuclear reactions from protons to heavy actinides
(\cite{Ter01}).
The  nuclear reaction network
includes  nuclei from beta-stability to the neutron-drip line.
For $10 \leq Z \leq 94$, we have used the network of Meyer et al.
(1992), which includes about $3000$ nuclear species.
We have extended this network to include lighter nuclei as well
as almost all charged-particle reactions for A$\leq 28$
(\cite{Kaj90}; \cite{Ori97}) plus all
$(\alpha,n)$ reactions up to Z$=36$.
For the nucleosynthesis calculation we have checked the results
with and without
$\nu_e$ capture on nuclei
 (\cite{Meyer98}).  The results do not
depend upon this  effect, presumably because material is so quickly
removed from the neutron star surface.

\subsection{$r$-Process Yields}
The ejected mass shells of Figure 1 were evolved with our $r$-process
network.  Figure 4 shows the results of several trajectories
labeled by their Lagrangian zone numbers.
We see that material
ejected at the bottom contribute to the heaviest $A = 195$ peak,
while those shells which had higher $Y_e$ only contribute to
the $A = 130$ peak and below.  Shells with $Y_e > 0.25$ only
produce iron group and elements up to $A = 100$.  They
do not contribute to the synthesis of $r$-process elements.

The nucleosynthesis yields of these different mass shells were summed
to produce the final $r$-process abundance curve shown
in Figure 5.
Here, we see an adequate reproduction of the Solar $r$-process
abundances.  The dips above and below  the $A \approx 130$ peak
in our computed abundance curve may
be an indication (cf.~\cite{Meyer92})
that the $N=82$ shell closure is too strong in our
mass formula (\cite{hilf}).  Other than that, the abundances of nuclei
heavier than $A \sim 100$ are nicely reproduced.

The fact that
elements with $A < 100$ are underproduced suggests an interesting possibility.
These are precisely the nuclei overproduced in the neutrino-driven
wind scenario.  They are also the elements which seem to deviate
from the Solar $r$-process abundances in metal-poor halo stars
(Ryan, Norris, \& Beers 1996; Sneden et al.~1996; 1998; 2000).
The suggestion is therefore  that
two different environments may  contribute to the  $r$-process:
one being the prompt explosion of low-mass SNe progenitors
under study here.  Such environments may
produce the heaviest $r$-process nuclei; the other environment could be
the neutrino-driven winds in more massive progenitors.
These events might produce
the lighter $A \le 100$ nuclides.

This is consistent with recent
meteoritic evidence (Qian  2000; Qian \& Wasserburg 2000)
that two different $r$-process
environments preceded the formation of the Solar System.
It is also consistent with  galactic
evolution evidence.   Ishimaru \& Wanajo (1999) have demonstrated
that the observed large dispersion of $r$-process elements in metal-poor
halo stars can be explained if the r-process
occurred in either low-mass (8-10 M$_\odot$)
or high mass ($\sim 30$ M$_\odot$) supernova progenitors.  Here we 
favor low-mass progenitors, but perhaps
both environments contribute.

\subsection{Galactic Contribution}
Regarding the $r$-process yields implied by the present study,
we see from Figure 3 that $\approx 0.025$ M$_\odot$ of $r$-process material
is ejected in our 11 M$_\odot$ model.  This corresponds to a mass fraction
in the
ejecta of $X_r(m=11) \approx 3 \times 10^{-3}$.   However, since the
prompt mechanism can only work in low mass progenitors, only supernovae
in a narrow mass range, say  $\Delta m \sim 1$ M$_\odot$
probably contribute to
the galactic abundances.  Hence, to estimate the average yield
per supernova, we must correct for the fraction of supernovae
which can contribute.  Assuming a standard Salpeter initial mass function
$\phi(m) \propto m^{-2.3}$, then the total average mass
fraction of ejected  $r$-process material
per supernova $\langle X_r \rangle_{SN}$ becomes:
\begin{eqnarray}
\langle X_r \rangle_{SN} &=& {X_r(m=11)
\times (m - m_{rem}) \phi(11) \Delta m
\over \int_{11}^\infty (m - m_{rem}) \phi(m)  dm} \nonumber \\
& \approx &
{0.3 X_r(m=11) \Delta m  \over 11~{\rm M}_\odot} \approx 7.5 \times 10^{-5}~~,
\end{eqnarray}
where in the above we have ignored the mass of the remnant compared to
the mass of ejecta.   Now assuming that the total SNII rate has been
constant at $10^{-2}$ yr$^{-1}$ over the past $10^{10}$ yr, then
we would estimate a current mass in $r$-process elements of
the Galaxy of $M_r^{Gal} \approx 7.5 \times 10^4$ M$_\odot$.
Assuming a total baryonic mass for the Galaxy of $10^{11}$ M$_\odot$,
then the present mass fraction in $r$ elements would be $\sim 7.5 \times
10^{-7}$  in reasonable agreement with the observed Solar
mass fraction of $\sim 10^{-7}$.

\section{Conclusion}
   We conclude that the prompt mechanism
  in low-mass  supernova progenitors
remains as a viable
model for the $r$-process.  
Clearly, one must still establish whether prompt explosions
of the sort described here actually occur when neutrino transport
is included.  Nevertheless, our studies show that if prompt
explosions occur, one 
objection to prompt $r$-process nucleosynthesis is avoided.
There is
no over-production of $r$-process elements as long as an
accurate accounting of the neutronization of the prompt ejecta
  is made.  It also provides
a very good reproduction of the Solar $r$-process abundance distribution
for elements with $A \ge 100$, suggesting that this might be the source
of the heaviest $r$-process nuclei while perhaps
delayed neutrino-driven winds or neutron-star mergers  may
be the source of lighter $r$-process nuclei.  
Obviously, a key component of this
paradigm is to understand just which progenitors are capable of producing
a prompt explosion and $r$-process nucleosynthesis and also
whether the late time neutrino heating will affect the
$r$-process yields.  In future work we
will address both of these points, however, regarding the delayed
neutrino emission we expect that it will not affect the results reported here
as by the time the delayed neutrinos arrive, the prompt ejecta is already
far from the proto-neutron star.

\acknowledgments

The authors wish to thank  J. Cowan, W. Hillebrandt and S. Wanajo
for useful discussions on the $r$-process.
We also thank  Stan Woosley for providing his numerical data of
presupernova models.  One of the authors (KS) would like to
acknowledge the collaborative efforts of H. Shen, K. Oyamatsu and H. Toki
for providing the relativistic EOS table.
Numerical simulations have been performed on the supercomputers
at RIKEN and KEK (KEK Supercomputer Project No. 00-63, 01-75).
One of the authors (MT) wishes to acknowledge fellowship support
from RIKEN  as a Junior Research Associate.
  This work was supported in part
by Japan Society for Promotion of Science, and by the Grant-in Aid for
Scientific Research (1064236, 10044103, 11127220, 12047230, 12740138)
of the Ministry of Education, Science, Sports and Culture of Japan.
Work at Univ. Notre Dame is  supported by
DoE Nuclear Theory Grant DE-FG02-95-ER40934.

\begin{figure*}[t]
\epsscale{1.3}
\plotone{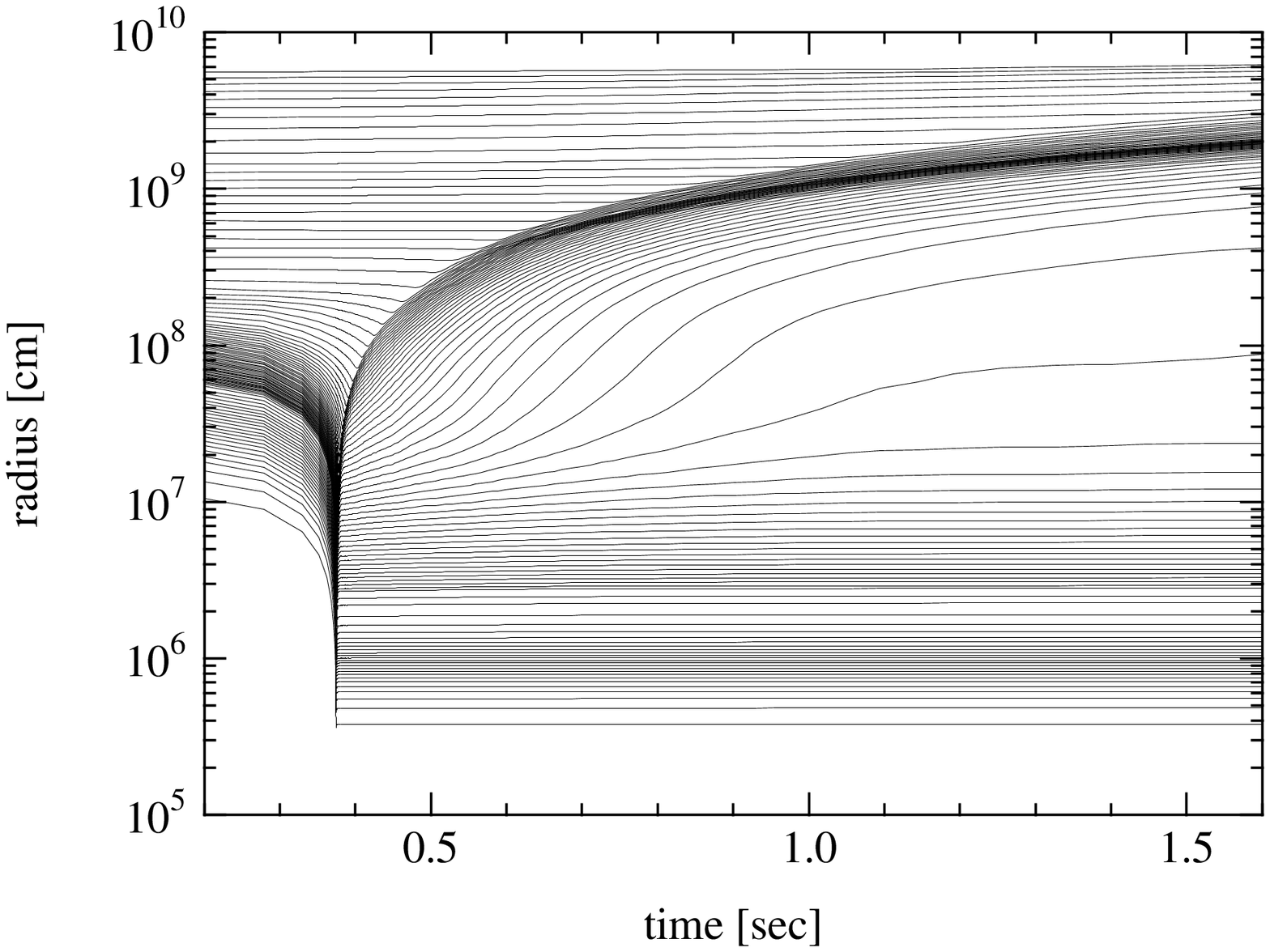}
\figcaption{Evolution of  the spatial coordinate
for various mass shells in the collapse
and prompt explosion of our 11 M$_\odot$ model.}
\label{fig:1}
\end{figure*}

\begin{figure*}[t]
\epsscale{1.0}
\plotone{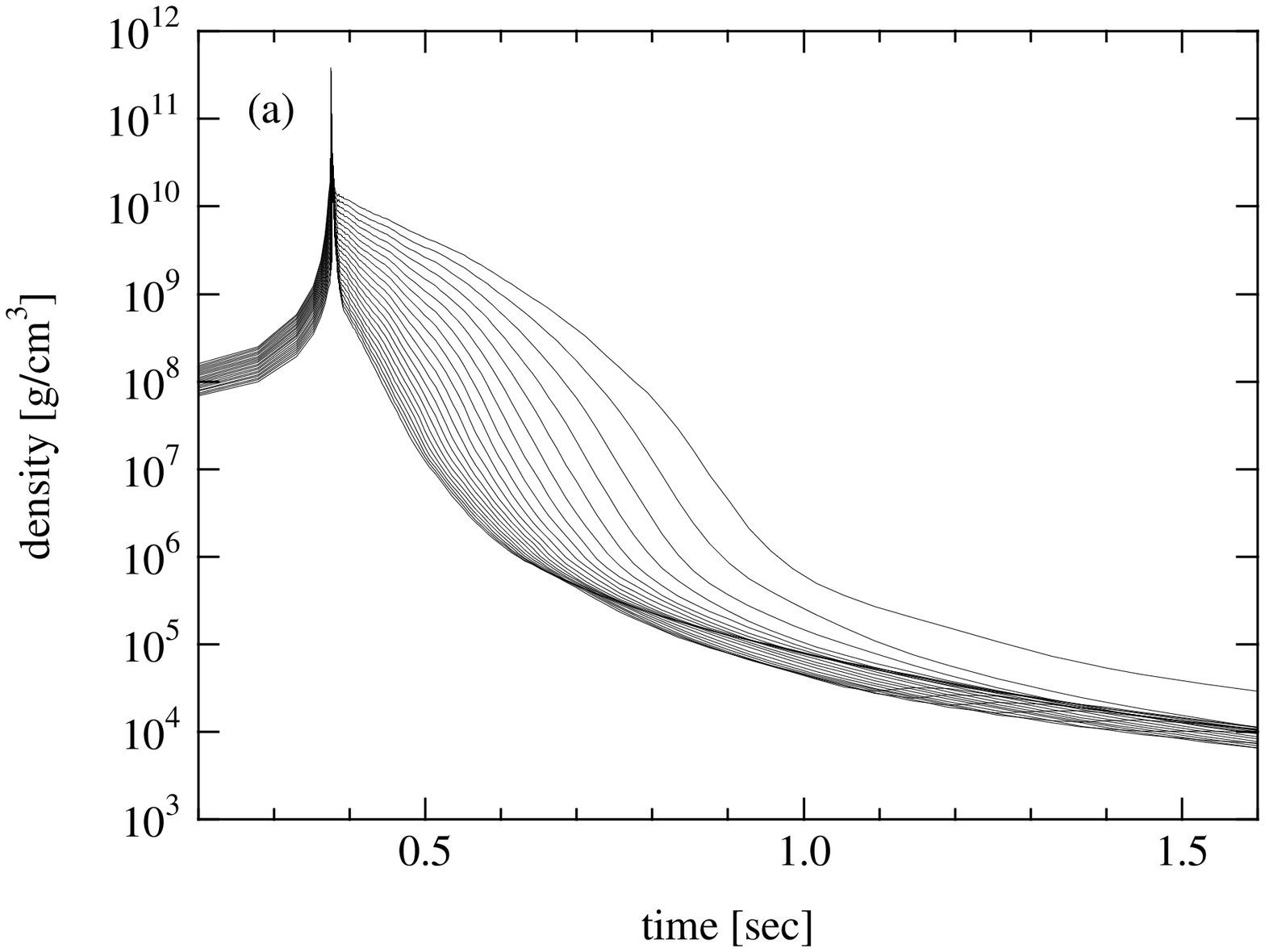}
\plotone{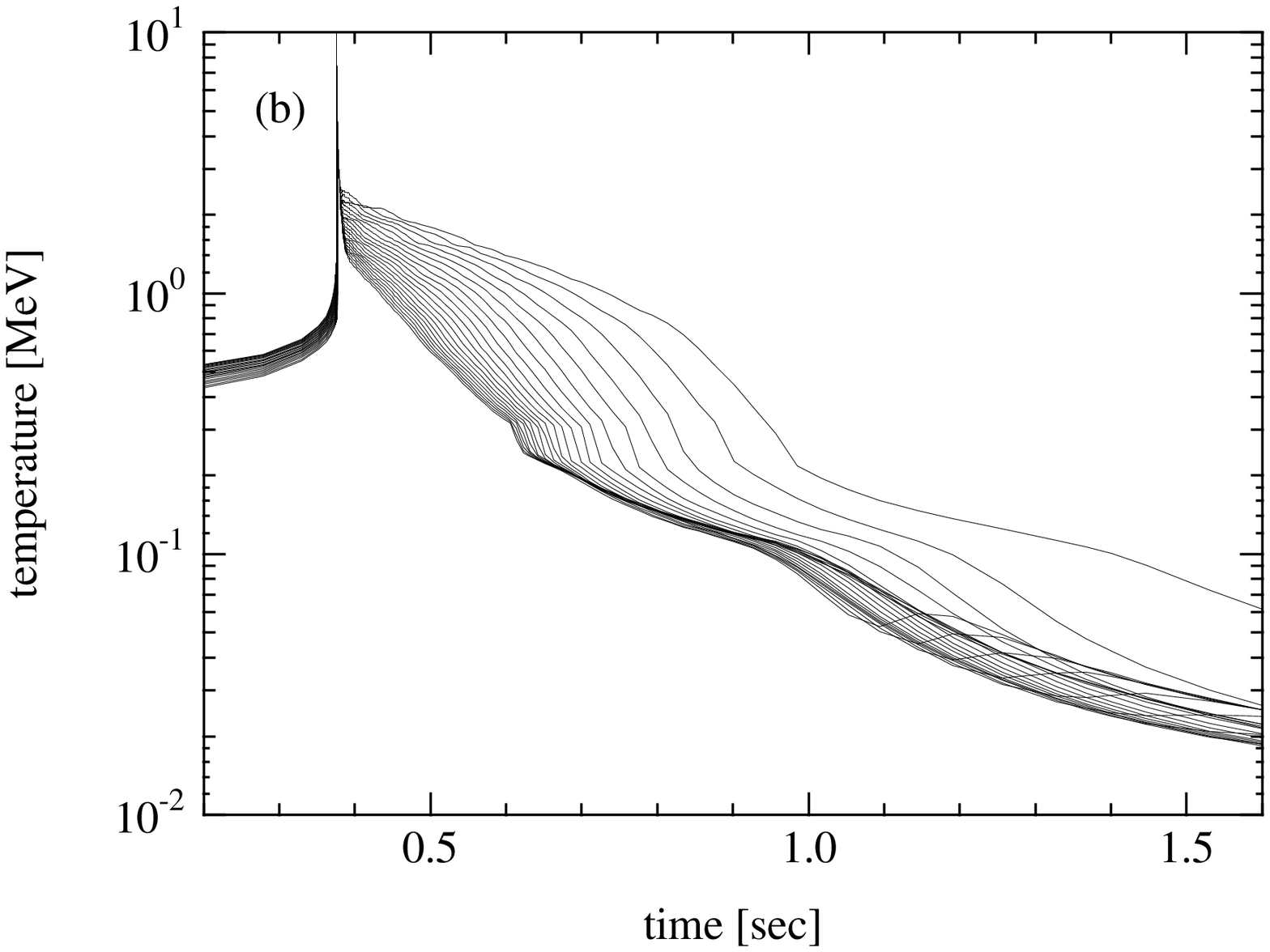}

\plotone{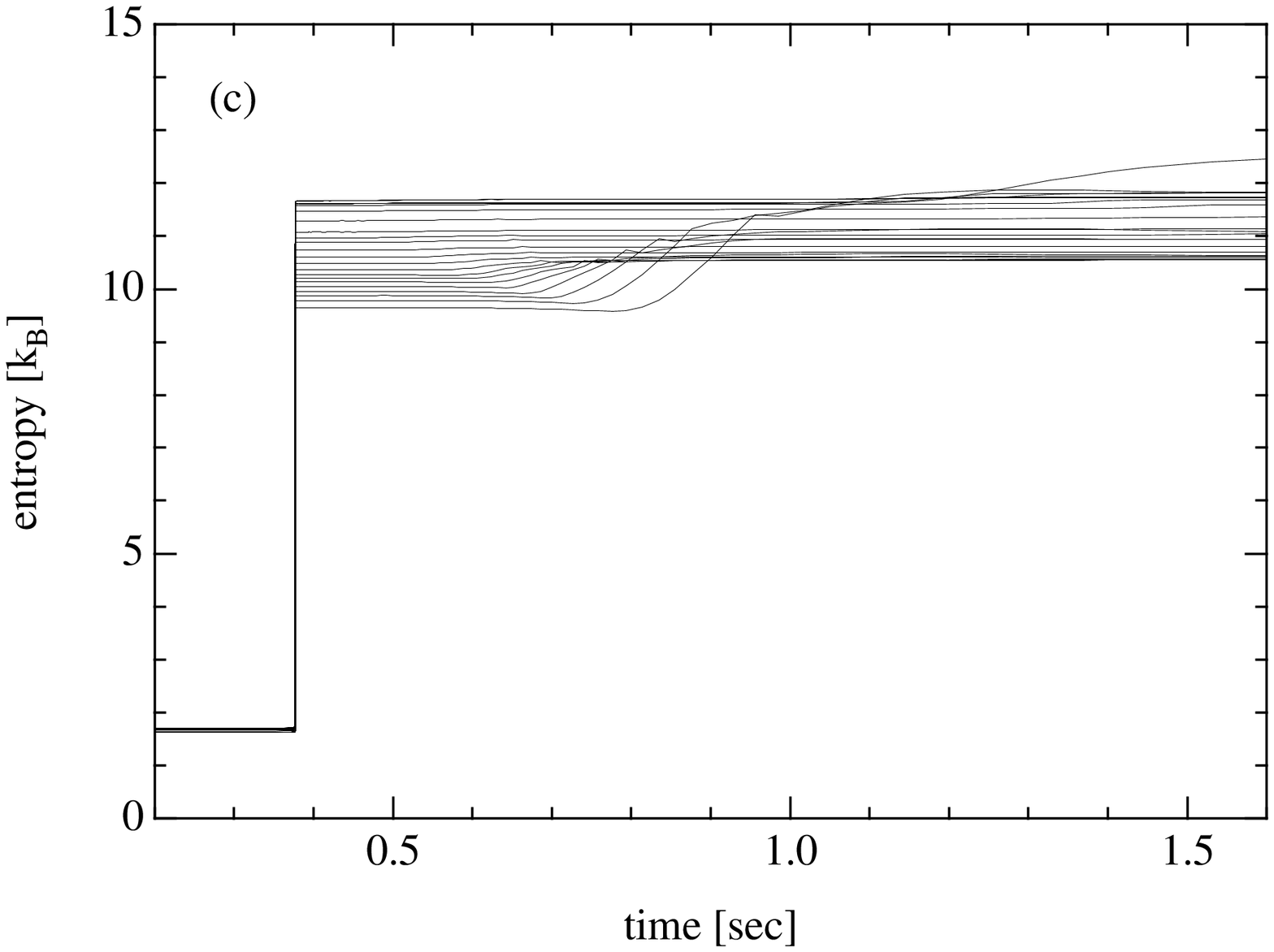}
\plotone{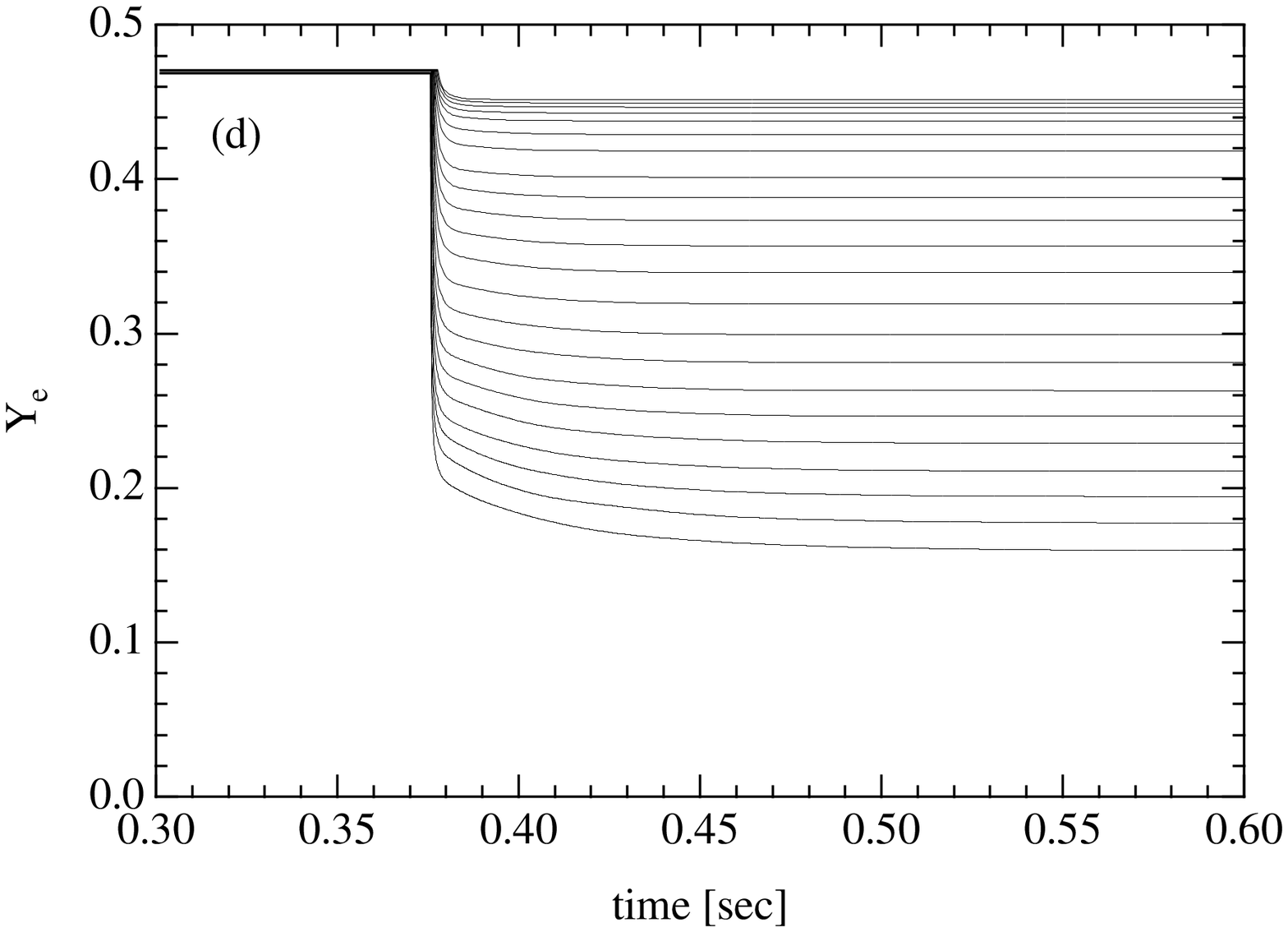}
\figcaption{Evolution of: (a) baryonic mass density; (b) temperature;
(c) entropy; and (d) electron fraction $Y_e$
as a function of time for ejected material in the prompt explosion.
Note that the time interval is expanded for (d) to better show
the rapid change in $Y_e$.}
\label{fig:2}
\end{figure*}

\begin{figure*}[t]
\epsscale{1.3}
\plotone{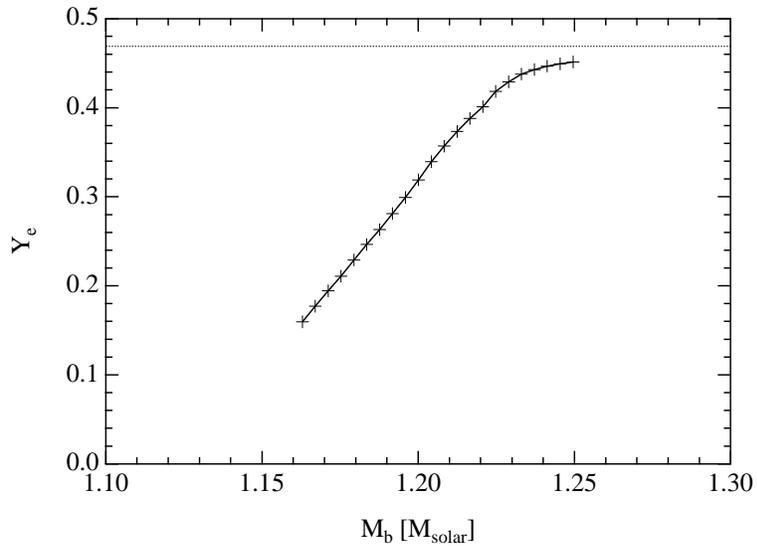}
\figcaption{Electron fraction $Y_e$ as a function of
baryon mass coordinate for ejected material in the prompt explosion.}
\label{fig:3}
\end{figure*}

\begin{figure*}[t]
\epsscale{1.3}
\plotone{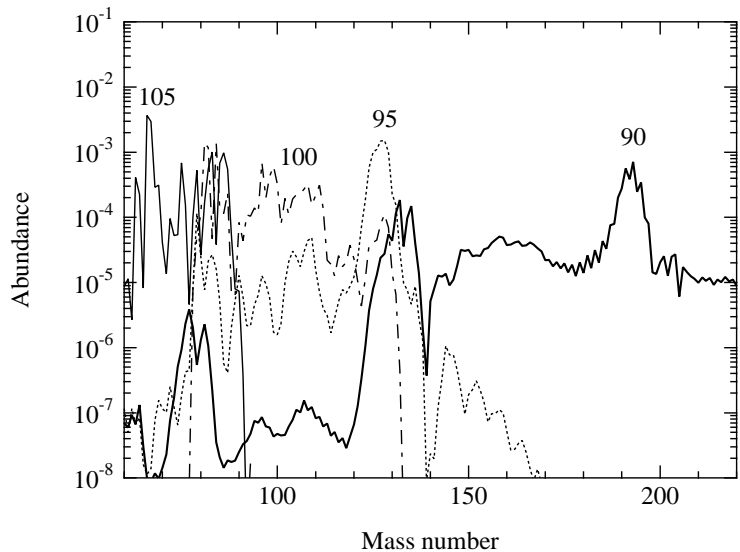}
\figcaption{Abundances computed for ejected
material from several representative mass shells:
90 ($Y_e = 0.16$);
95 ($Y_e = 0.25$);
100 ($Y_e = 0.34$);
105 ($Y_e = 0.42$).}
\label{fig:4}
\end{figure*}

\begin{figure*}[t]
\epsscale{1.3}
\plotone{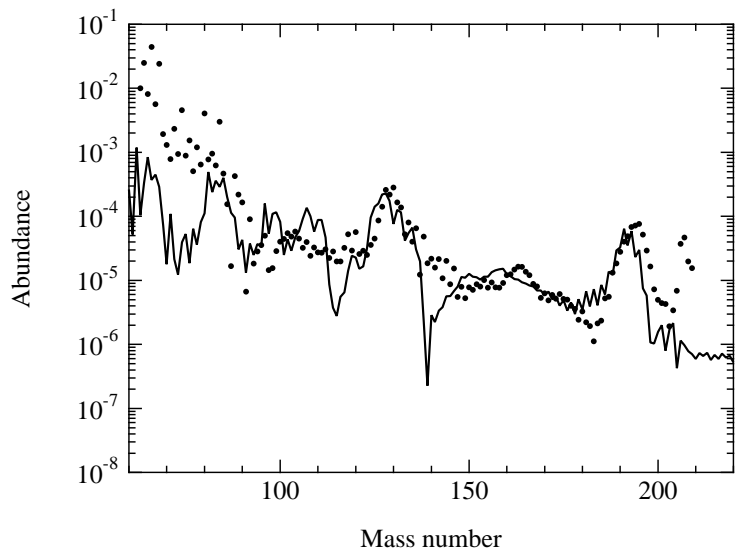}
\figcaption{Final integrated isotopic abundances for ejected
material (solid line) compared with the Solar $r$-process abundances
(filled circles; \cite{Kapp89}).}
\label{fig:5}
\end{figure*}

\end{document}